\documentclass[conference]{IEEEtran}
\IEEEoverridecommandlockouts

\usepackage{cite}
\usepackage{amsmath,amssymb,amsfonts}
\usepackage{algorithmic}
\usepackage{graphicx}
\usepackage{textcomp}
\usepackage{xcolor}
\usepackage{listings}
\usepackage{xurl}  
\usepackage{hyperref} 
\usepackage{xspace}
\usepackage{comment}
\usepackage[paper=letterpaper,
            left=0.75in,   
            right=0.75in,  
            bottom=0.75in, 
            top=0.75in        
           ]{geometry}

\newcommand{\qq}[1]{``#1''}

\newcommand{\LLM}{LLM\xspace}
\newcommand{\LLMs}{LLMs\xspace}

\newcommand{\eg}{e.g.,\xspace}
\newcommand{\ie}{i.e.,\xspace}
\newcommand{\FtwoScore}{\ensuremath{\text{$F_{2}$-score}}\xspace}
\newcommand{\TPR}{\ensuremath{\text{TPR}}\xspace}
\newcommand{\FPR}{\ensuremath{\text{FPR}}\xspace}
\newcommand{\Precision}{\ensuremath{\text{Precision}}\xspace}
\newcommand{\highlightTerm}[1]{\emph{\textbf{#1}}}
\usepackage[angle=0,pos={0.5\paperwidth,1cm},hanchor=c,vanchor=m]{draftwatermark}
\SetWatermarkText{Preprint of paper at the 2025 IEEE International Conference\\on Intelligence and Security Informatics (IEEE ISI 2025)}
\SetWatermarkScale{0.125}

\begin{document}

\title{Towards Effective Complementary Security Analysis using Large Language Models%
}

\newcommand{\mybox}[1]{\parbox[t]{6.75cm}{\centering#1}}
\newcommand{\myboxB}[1]{\parbox[t]{2cm}{\centering#1}}
\newcommand{\abox}[3]{\mybox{#1}\myboxB{#2}\mybox{#3}}
\newcommand{\shift}{\parbox[t]{1.5cm}{\ }}

 \author{
 \IEEEauthorblockN{
     Jonas Wagner\IEEEauthorrefmark{1}\IEEEauthorrefmark{2}, 
     Simon Müller\IEEEauthorrefmark{3},
     Christian Näther\IEEEauthorrefmark{3},
     Jan-Philipp Steghöfer\IEEEauthorrefmark{3},
     Andreas Both\IEEEauthorrefmark{2}
 }
  	\IEEEauthorblockA{
         \IEEEauthorrefmark{1}\textit{Data Science Group}, \textit{Paderborn University}, Paderborn, Germany\\
         \IEEEauthorrefmark{2}\includegraphics[height=2ex]{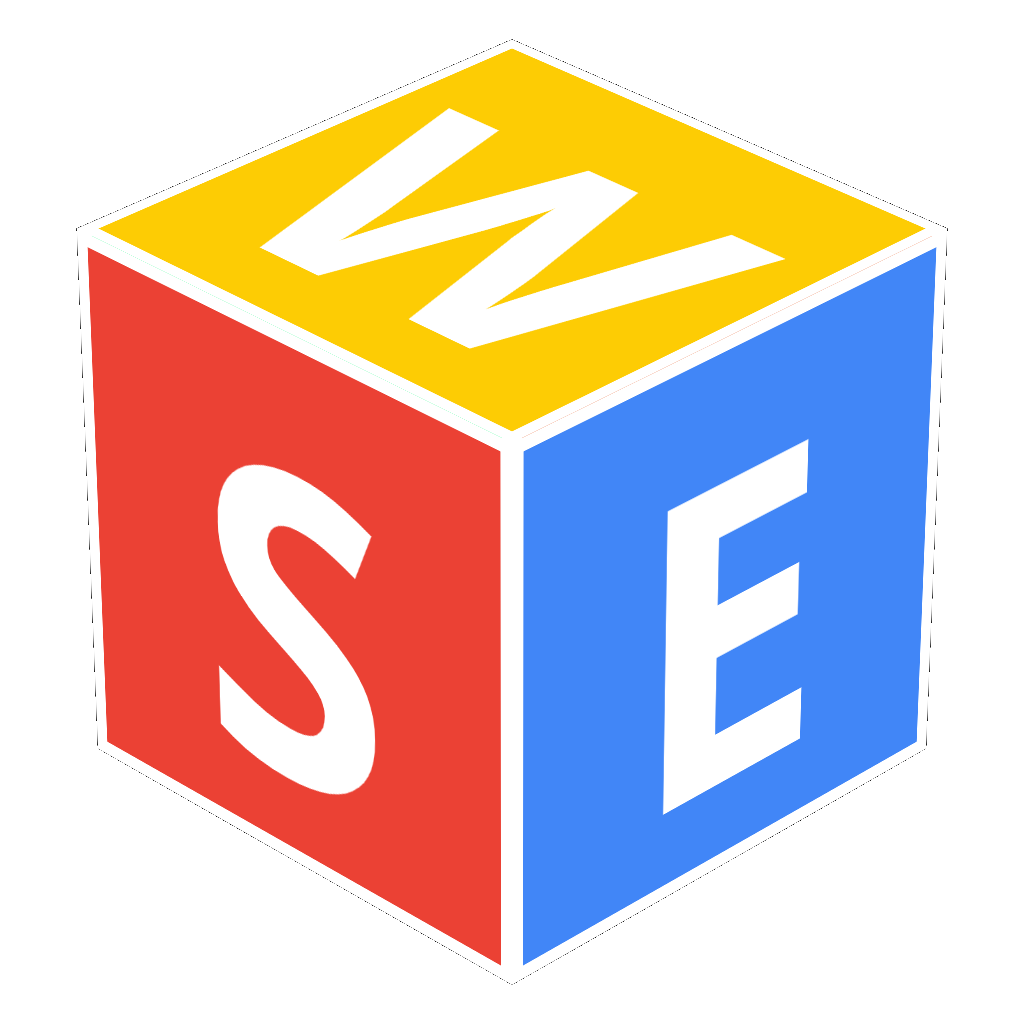}\,\textit{Web \& Software Engineering Research Group}, \textit{Leipzig University of Applied Sciences}, Leipzig, Germany\\
         \IEEEauthorrefmark{3}\textit{XITASO GmbH IT \& Software Solutions}, Augsburg, Germany\\
         {jonas.wagner@uni-paderborn.de,} {\{simon.mueller, christian.naether, jan-philipp.steghoefer\}@xitaso.com,}\\ {\{jonas.wagner,andreas.both\}@htwk-leipzig.de}
  	}
}

\maketitle

\begin{abstract}
A key challenge in security analysis is the manual evaluation of potential security weaknesses generated by static application security testing (SAST) tools.
Numerous false positives (FPs) in these reports reduce the effectiveness of security analysis.
We propose using Large Language Models (\LLMs) to improve the assessment of SAST findings. 
We investigate the ability of LLMs to reduce FPs while trying to maintain a perfect true positive rate, using datasets extracted from the OWASP Benchmark (v1.2) and a real-world software project.
Our results indicate that advanced prompting techniques, such as Chain-of-Thought and Self-Consistency, substantially improve FP detection. 
Notably, some \LLMs identified approximately 62.5\% of FPs in the OWASP Benchmark dataset without missing genuine weaknesses. Combining detections from different \LLMs would increase this FP detection to approximately 78.9\%.
Additionally, we demonstrate our approach's generalizability using a real-world dataset covering five SAST tools, three programming languages, and infrastructure files. The best \LLM detected 33.85\% of all FPs without missing genuine weaknesses, while combining detections from different \LLMs would increase this detection to 38.46\%. Our findings highlight the potential of \LLMs to complement traditional SAST tools, enhancing automation and reducing resources spent addressing false alarms.
\end{abstract}


\vspace*{-0.15ex}
\section{Introduction}

Ensuring software security involves identifying weaknesses and addressing potential vulnerabilities early in the development process. 
While static application security testing (SAST) tools are essential for detecting potential issues, they produce a large number of incorrectly reported vulnerabilities, often between 30--100\% \cite{MitigatingFalsePositiveStaticAnalysisWarnings2023}, requiring time-consuming manual reviews by security experts\cite{MitigatingFalsePositiveStaticAnalysisWarnings2023,Falsenegativethatoneisgoingtokillyou2024}. 
This inefficiency can lead to missed threats or wasted resources, making it crucial to enhance the accuracy of SAST tools.
However, developers \qq{want tools to detect real vulnerabilities}\ \cite{Falsenegativethatoneisgoingtokillyou2024}.
Large Language Models (\LLMs) have emerged as a promising solution to address these challenges. 
By leveraging their advanced capabilities to analyze code, \LLMs offer the potential to significantly reduce false positives (FPs) in security assessments.
This study investigates the integration of \LLMs with SAST tools to optimize security evaluations. 
Specifically, we explore how advanced prompting techniques, such as Chain-of-Thought (CoT) reasoning and Self-Consistency (SC), can be used to enhance the reassessment of security findings generated by SAST tools.
A key objective of our research is the development of a generalized assessment strategy designed to identify FPs in security reports while trying to maintain a perfect true positive (TP) rate (\TPR). 
We term this approach \highlightTerm{conservative analysis}. 
In the context of security analysis, missing even a single genuine weakness can lead to a critical vulnerability, potentially compromising an entire system and thereby invalidating the approach. 
Consequently, we investigate the effectiveness of \LLMs in implementing this conservative analysis approach, specifically assessing their capability to reliably detect FPs within datasets of security findings while ensuring that no genuine weaknesses are missed.
To guide this investigation, we focus on the following research questions:
\begin{description}
    \item[\textbf{RQ1}] Are \LLMs capable of detecting incorrectly reported security findings (FPs), without missing any genuine weakness (TPs)?
    \item[\textbf{RQ2}] Assuming a successful conservative analysis (\ie \quad $\text{TPR} = 100\%$), how effectively can \LLMs detect FPs ($\text{Effectiveness} = \frac{\text{detected FP}}{\text{existing FPs}}$)?
    \item[\textbf{RQ3}] To what extent can combining the predictions of multiple \LLMs enhance the effectiveness of conservative FP detection?
\end{description}
By addressing these questions, this research contributes to the ongoing evolution of automated cybersecurity workflows. 
It highlights the potential of \LLMs to enhance the effectiveness and accuracy of security assessments without the need for resource-intensive fine-tuning, thereby aligning with the rapid advancements of general-purpose foundation \LLMs.

\section{Background and Related Work}
SAST tools are fundamental for identifying weaknesses within software by analyzing source code without execution.
They typically approximate program behavior via control-flow graphs and abstract syntax trees, then match known vulnerability patterns~\cite{spa}.
Despite their value, they frequently produce FPs, requiring costly manual reviews \cite{MitigatingFalsePositiveStaticAnalysisWarnings2023,Falsenegativethatoneisgoingtokillyou2024,spa}. 
They often categorize findings based on Common Weakness Enumerations (CWEs), emphasizing coding flaws rather than specific exploits.
The OWASP Benchmark~\cite{owasp_benchmark} offers a standardized resource to evaluate the effectiveness of these tools across various CWE categories and has been used extensively in evaluating SAST tools\cite{MitigatingFalsePositiveStaticAnalysisWarnings2023,koc_learning_2017,koc_empirical_2019}. 

Recently, researchers have investigated the use of \LLMs for direct weakness and vulnerability detection, demonstrating that models such as GPT variants can match or even outperform conventional SAST tools~\cite{thapa_transformer-based_2022,bakhshandeh_using_2023,jensen_software_2024}.
Bakhshandeh et al.~successfully combined ChatGPT with tools like Bandit and Semgrep, though their experiments were confined to Python code~\cite{bakhshandeh_using_2023}.
Wen et al.~introduced the \textsc{Llm4Sa} method for automated inspection of static bug warnings across C/C++ benchmarks, using advanced prompting strategies like CoT and few-shot learning, although their approach occasionally missed genuine vulnerabilities~\cite{wen_automatically_2024}.
Li et al.~applied ChatGPT to summarize functions and reassess Linux kernel security findings, effectively identifying FPs, yet their work was constrained by small-scale experiments focused on use-before-initialization bugs~\cite{li_assisting_2023}. 
Zhou et al.~compared multiple SAST tools and \LLMs across several languages, finding that \LLMs had higher detection rates but produced more FPs and did not adopt conservative analysis strategies that prevent missing genuine weaknesses~\cite{zhou_comparison_2024}. 
Similarly, Tamberg et al.~benchmarked various prompting techniques and proprietary \LLMs against traditional tools like CodeQL and SpotBugs, observing increased vulnerability recall but higher FP rates from \LLMs; however, their evaluation lacked verification against diverse real-world datasets~\cite{tamberg_harnessing_2025}.

Fine-tuning \LLMs on security datasets can enhance their effectiveness but typically requires significant computational resources~\cite{thapa_transformer-based_2022}. 
Recent advances in foundation \LLMs, such as GPT-4, have made fine-tuning less critical, demonstrating state-of-the-art performance through prompt engineering alone~\cite{nori_can_2023}. 
Our research leverages these advances, exploring how foundation \LLMs can complement SAST tools by specifically targeting FP reduction and improving the effectiveness of security assessments, rather than focusing solely on direct vulnerability detection.
Effective use of \LLMs relies heavily on advanced prompting techniques, including zero-shot and few-shot prompting, CoT reasoning, and SC prompting~\cite{wei_chain--thought_2023,kojima_large_2023,wang_self-consistency_2023}. 
However, prior research has yet to comprehensively investigate how these prompting strategies can specifically improve the assessment of SAST findings and which contextual information is most impactful for accurate FP detection.

Existing research often suffers from limited or homogeneous datasets~\cite{bakhshandeh_using_2023,li_assisting_2023,wen_automatically_2024,tamberg_harnessing_2025}, a lack of conservative analysis approaches (\ie risk of missing a genuine weakness)~\cite{wen_automatically_2024,zhou_comparison_2024} and reliance primarily on proprietary models~\cite{bakhshandeh_using_2023,li_assisting_2023,tamberg_harnessing_2025}, restricting practical deployment in privacy-sensitive scenarios.
To address these limitations, our work explicitly adopts a conservative analysis approach, ensuring genuine weaknesses are not overlooked.
Crucially, we integrate both proprietary and open-source \LLMs to facilitate private deployments and secure analyses, uniquely extending applicability from traditional software code to infrastructure code. 
Furthermore, we employ both a benchmark dataset for initial experiments and a real-world dataset of security findings to underscore the practical relevance of our approach within real software projects.
Through these contributions, our methodology significantly advances current research by providing a generalized, conservative, and broadly applicable approach that leverages the complementary strengths of \LLMs and SAST tools.

\section{Concept}\label{sec:approach}
\begin{figure*}[t]
  \centering
  \includegraphics[width=1\linewidth]{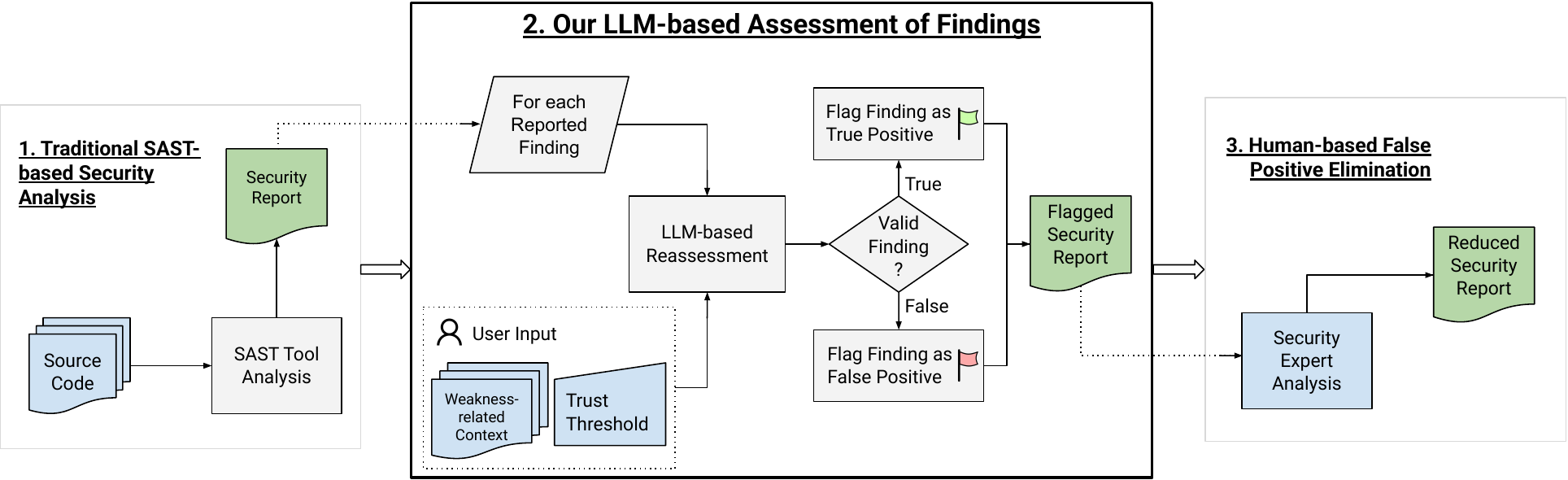}
  \caption{Complementary Security Analysis Process}
  \label{fig:concept}
\end{figure*}

Our objective is to develop a comprehensive and generalizable framework that leverages the strengths of both \LLMs and SAST tools to enhance the effectiveness of static security assessments. 
As depicted in Fig.~\ref{fig:concept}, our approach extends traditional SAST-based security analyses by integrating an additional \LLM-based assessment stage prior to the final human validation step.

\subsection{Stage 1: Traditional SAST-based Security Analysis}
The first stage, represented by the box on the left side of Fig.~\ref{fig:concept}, illustrates a traditional SAST workflow.
In this stage, the source code is analyzed by a SAST tool, generating a security report. 
Each finding in this report highlights a potential weakness within the code. 
However, as noted in prior research, SAST tools typically produce numerous FPs, necessitating substantial manual effort from security experts to validate each finding. This significantly reduces the effectiveness of the security assessment process.

\subsection{Stage 2: Our LLM-based Assessment}
To tackle the resource-intensive process of detecting FPs, we introduce an \LLM-based assessment, depicted in the central box of Fig.~\ref{fig:concept}.
For each code fragment, our framework generates a tailored prompt that includes the pertinent source code, essential contextual information (such as the CWE-ID or line of code), and explicit instructions for the \LLM to produce a structured reasoning process. 
Instead of returning a simple binary decision, the \LLM is instructed to provide an in-depth reasoning process along with a numerical score that reflects its degree of agreement with the SAST tool's classification. 
Based on a user-defined threshold, each finding is then flagged as either a TP (genuine security weakness) or a FP (secure code wrongly reported by the SAST tool), resulting in a flagged security report.

\subsection{Stage 3: Human-Based False Positive Elimination}
The flagged security report is then provided to security experts, as illustrated on the right side of Fig.~\ref{fig:concept}. 
Given our objective to perform a conservative analysis of the report{}’s security findings, security experts could then utilize this flagged report to further analyze findings that were flagged as TPs. Since the analysis conducted by the \LLM is assumed to be conservative, findings flagged as FPs could be ignored, which would significantly reduce the resource expenditure associated with manual security reviews.

\subsection{Datasets}\label{concept_dataset}
We employed two distinct datasets in our work. 
First, we generated a security report using the OWASP Benchmark (v1.2)~\cite{owasp_benchmark}, analyzed with SpotBugs using the FindSecBugs plugin, resulting in 2,015 security findings. 
This dataset was randomly split into a training subset (80\%, 1,557 samples) and a testing subset (20\%, 403 samples—275 TPs and 128 FPs). 
Both subsets include findings from eleven distinct vulnerability categories, each identified by unique CWE-IDs (\eg CWE-501: Trust Boundary Violation)~\cite{wagner_2025_15378450}.
To demonstrate that our generalized framework not only improves security analysis on benchmark datasets, which might implicitly be part of a model's training data, but also performs effectively in real-world settings, we created a second dataset composed of real-world security findings.
This dataset includes findings generated by applying five different SAST tools (Checkov, CodeQL, Semgrep, KICS, and SpotBugs with the FindSecBugs plugin) to the partly open-source software project \emph{Mnestix}\footnote{\href{https://github.com/mnestix}{https://github.com/mnestix}}, developed by XITASO GmbH.
It covers a diverse set of programming languages and infrastructure file types, including Java, C\#, TypeScript, and Dockerfiles.
A panel of three senior security experts, selected based on their familiarity with static code analysis, manually assessed 114 security findings (49 TPs and 65 FPs), thereby establishing a reliable ground truth~\cite{wagner_2025_15378450}.

\subsubsection*{Evaluation}\label{concept_evaluation}
To evaluate our approach, we measure the effectiveness on different datasets using common classification metrics.
These include the \TPR, defined as $\TPR = \frac{TP}{TP + FN}$; the \FPR, defined as $\FPR = \frac{FP}{FP + TN}$; and \Precision, defined as $\Precision = \frac{TP}{TP + FP}$.
Regarding \textbf{RQ1}, our primary focus is on achieving a conservative analysis.
To emphasize this goal, we introduce a weighted F-score defined as:
$\FtwoScore = \frac{(1 + \beta^2) \times P \times R}{(\beta^2 \times P) + R}$
with \(\beta\) controlling the weight given to recall over precision. 
Following the example of Christen, Hand, and Kirielle, we chose \(\beta = 2\), which emphasizes recall~\cite{10.1145/3606367}.
On the OWASP Benchmark datasets, these metrics are computed per vulnerability category and then averaged to avoid bias from categories with more test cases.
In addition, the ratio of true negatives (TN) and false negatives (FN) of the \LLM's assessment classification is of special interest to us. 
A TN classification result of the \LLM's assessment means the \LLM successfully detected an FP of the SAST tool's security report, which should be maximized. 
FN classifications of the \LLM's assessment should be minimized, or at best kept at zero, as it describes cases where a genuine weakness that was detected by the SAST tool is mistakenly labeled as secure by the \LLM.
By integrating these metrics with a robust prompting strategy across a diverse set of \LLMs, our work aims not only to identify the \LLMs that best complement SAST tools but also to elucidate systematic approaches for guiding \LLM reasoning to optimize the security assessment process.

\section{Experiment Preparation and Setup}\label{sec:experiments}
\newcommand{\myparagraph}[1]{\textbf{#1:}\xspace}
\myparagraph{Preliminary Study}\label{exp_prelim_study}
In our preliminary study we used \texttt{GPT-3.5 Turbo} on the train split (1,557 samples), to first compare the importance of different contextual information generated by the SAST tool and additional CWE-related information, which we added, identifying the most important ones for our use case.
For a detailed overview of the used datasets, prompts and procedures of the preliminary study, refer to~\cite{wagner_2025_15378450}.
Contrary to our initial assumptions, our analysis indicates that the most beneficial contextual information is those supplied directly by the SAST tool, and integrating additional (\eg CWE-related) information did not produce measurable improvements.
The identified contextual information (produced by SpotBugs) include the weakness category, CWE-ID, method name, line of code, security risk type, and the complete source code file where the weakness was reported~\cite{wagner_2025_15378450}.
A plausible explanation could be that the \LLM employed for context comparison (\texttt{GPT-3.5 Turbo}) may already have internalized the relevant CWE-specific knowledge during its training, rendering additional CWE data superfluous.
Based on those results, in a second experiment of our preliminary study, we used the contextual information identified as most valuable to compare different prompting techniques, again on the train split (1,577 items) and using \texttt{GPT-3.5 Turbo}~\cite{wagner_2025_15378450}.
Our experiment compared a default prompt template in 0-shot, 3-shot, and 5-shot settings with a CoT approach also implemented in a 0-shot, 3-shot, and 5-shot setting. 
A performance gap emerged between the 0-shot and few-shot configurations, with CoT prompting being the most effective. Further, the 3-shot CoT strategy consistently matched the performance of 5-shot prompting, while at the same time requiring approximately 35\% fewer prompt tokens~\cite{wagner_2025_15378450}. 
For that reason, we decided to continue using 3-shot over 5-shot CoT prompting in this work.
Moreover, we observed that applying 3-shot CoT prompting five times repeatedly (following the SC approach) further enhanced performance compared to solely using CoT prompting; however, this improvement comes at the expense of five times the resource demands, including quintupled time, cost, and energy consumption.

\myparagraph{Experimental Setup}\label{exp_exp_setup}
To evaluate the performance differences between proprietary and open-source \LLMs, we conducted experiments employing 3-shot CoT prompting, leveraging the contextual information identified in our preliminary study. 
We specifically assessed each \LLM's ability to accurately detect FPs while maintaining a conservative analysis on the OWASP Benchmark's test split comprising 403 test cases. 
Additionally, we explored whether repeating 3-shot CoT prompting with SC across five iterations could further enhance detection accuracy. 
Finally, to robustly validate the generalizability and effectiveness of our framework, we evaluated the potential improvement gained from aggregating results from multiple \LLMs that successfully performed conservative analyses and further used the best-performing \LLMs to assess the security findings of our real-world dataset.
\subsubsection{Prompting Strategy} 
\begin{figure}[t]
\centering
\begin{minipage}{0.95\linewidth} 
\begin{lstlisting}[basicstyle=\footnotesize\ttfamily, frame=single,breaklines=true,breakatwhitespace=true]
<CoT few-shot examples>
Analyze the following potential vulnerability that was found by the security scanner "SpotBugs" with the "FindSecurityBugs"-Plugin in a Java source code project.
Vulnerability identified by the security scanner and contextual information: {context_items}
The source code included might be a false positive classification by the SAST scanner. Do you agree with the scanner that this source code contains an actual vulnerability?
Return a number for your decision ranging from "0.0" to "10.0", where "10.0" means you absolutely agree with the scanner, "0.0" means you absolutely do not agree, and any number around "5.0" means that you are not sure. Think step by step.
Give your answer in the following format: Explanation: "Let's think step by step..." Decision: 0.0 - 10.0
Explanation: 
\end{lstlisting}
\end{minipage}
\caption{Chain-of-Thought Prompt Template}
\label{fig:prompt}
\end{figure}

Each SAST finding and its corresponding code snippet are provided as input to the selected LLM, along with instructions to reason step-by-step about whether the reported weakness is valid or an FP. 
Rather than simply responding with \qq{TP} or \qq{FP,} the \LLM is instructed to (a) present a structured reasoning chain; and (b) conclude with a numeric score between \texttt{0.0} and \texttt{10.0}, where \texttt{0.0} indicates high confidence in an FP, \texttt{10.0} indicates high confidence in a true weakness, and \texttt{5.0} indicates uncertainty.
Fig.~\ref{fig:prompt} presents the prompt template that was crafted during our preliminary study and is subsequently used throughout our experiments.


\begin{figure*}[t]
  \centering
  \includegraphics[width=1\linewidth]{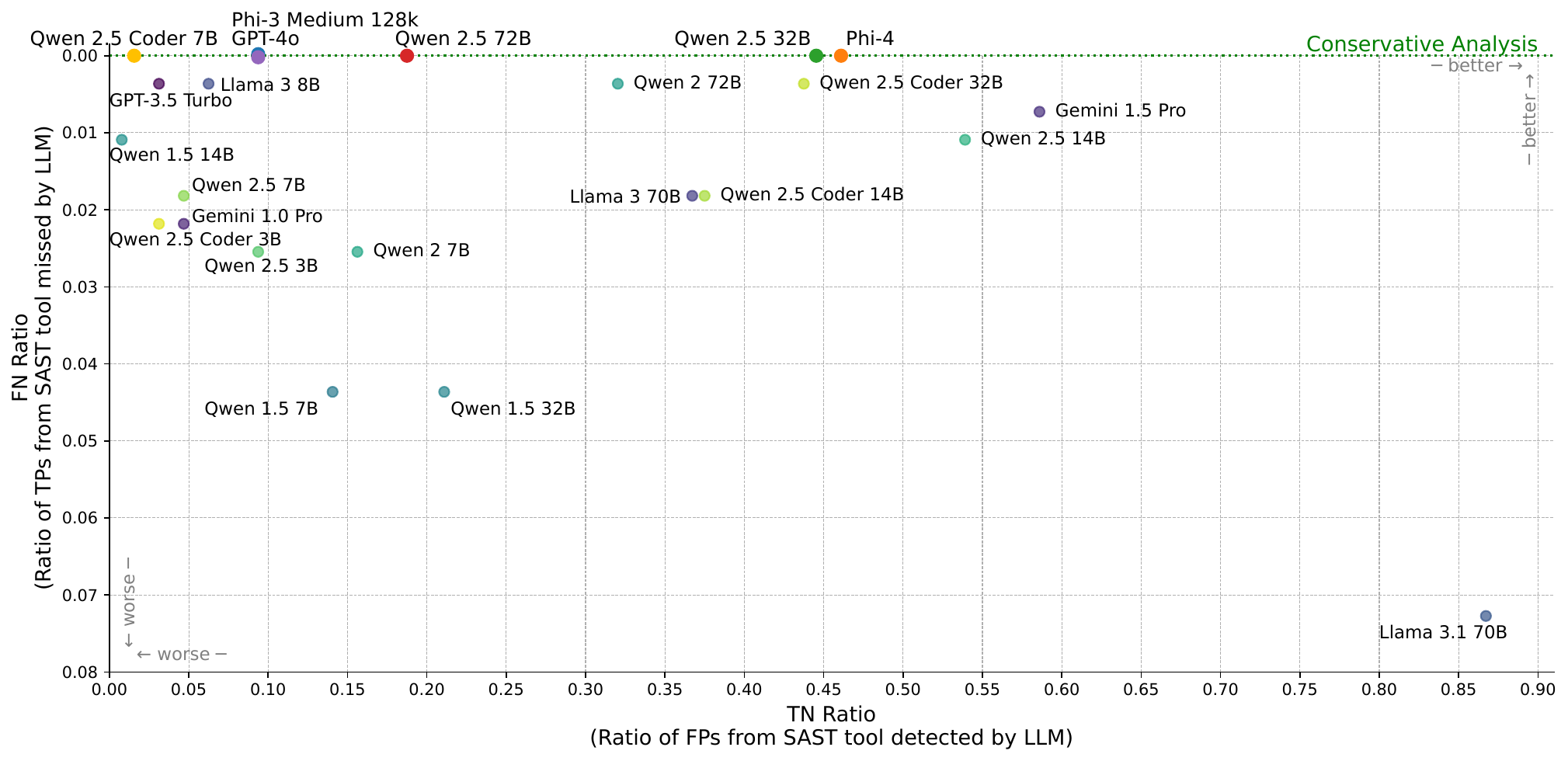}
  \caption{True Negative and False Negative Ratio of LLM-based Assessments at Threshold = 1 (y-axis inverted)}
  \vspace*{-1.5ex}
  \label{fig:llms_min_thershold}
  \vspace*{-0.5ex}
\end{figure*}

\begin{figure}[t]
  \centering
  \includegraphics[width=1\linewidth]{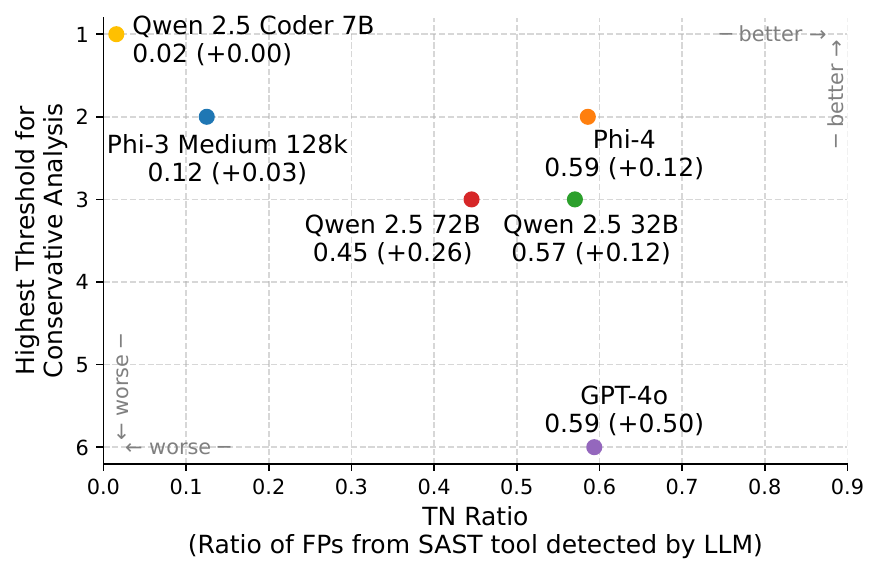}
  \vspace*{-1ex}
  \caption{True Negative Ratio of LLM Assessments across all Thresholds ensuring a Conservative Analysis (y-axis inverted)}
  \label{fig:llms_thresholds_comp}
  \vspace*{-1ex}
\end{figure}
\subsubsection{\LLM Selection and Parameters} 
To assess the generalizability of our prompting strategy, we evaluate a diverse set of open-source and proprietary \LLMs, including \texttt{GPT-4o}~\cite{openai_gpt-4_2024,gpt_4o}, Llama 3 (including \texttt{Meta-Llama-3-*-Instruct} and \texttt{Llama-3.1-*-Instruct} variants)~\cite{dubey_llama_2024}, \texttt{Gemini-*-Pro}~\cite{gemini_team_gemini_2024}, \texttt{Phi-3-Medium}~\cite{abdin_phi-3_2024}, \texttt{Phi-4}~\cite{abdin_phi-4_2024}, and a diverse subset of the Qwen model family (including \texttt{Qwen1.5-*-Instructs}, \texttt{Qwen2-*-Instructs}, \texttt{Qwen2.5-Coder-*-Instructs}, and \texttt{Qwen2.5-*-\linebreak[4]{}Instructs} variants)~\cite{bai_qwen_2023,yang_qwen2_2024,qwen_qwen25_2025,hui_qwen25-coder_2024}. 
This mix covers a very wide range of \LLM sizes, architectures, and training data scales.
To ensure comparability, we operate all \LLMs in this study with a context window of up to 8,192 tokens, accommodating both the provided code snippets and the few-shot CoT examples. 
We primarily rely on 3-shot CoT prompts, paired with an SC mechanism that regenerates the \LLM{}’s chain of thought using a higher temperature (\texttt{temperature=0.7}) to ensure robust exploration of reasoning paths. 
For the main inference run, we fix \texttt{temperature} at \texttt{0.0} to achieve outputs that are as deterministic as technically possible, minimizing randomness in the \LLM{}’s responses. 
Whenever supported, we prepend a system message that reads: \textit{You are a software security expert. Your main task is to analyze potential software vulnerabilities.}
This system-level instruction further emphasizes the security analysis context. 
By standardizing parameters across all \LLMs, we aim for a fair comparison of how different \LLM training procedures respond to our specific prompting strategy.

\begin{figure*}[t]
  \centering
  \includegraphics[width=0.92\linewidth]{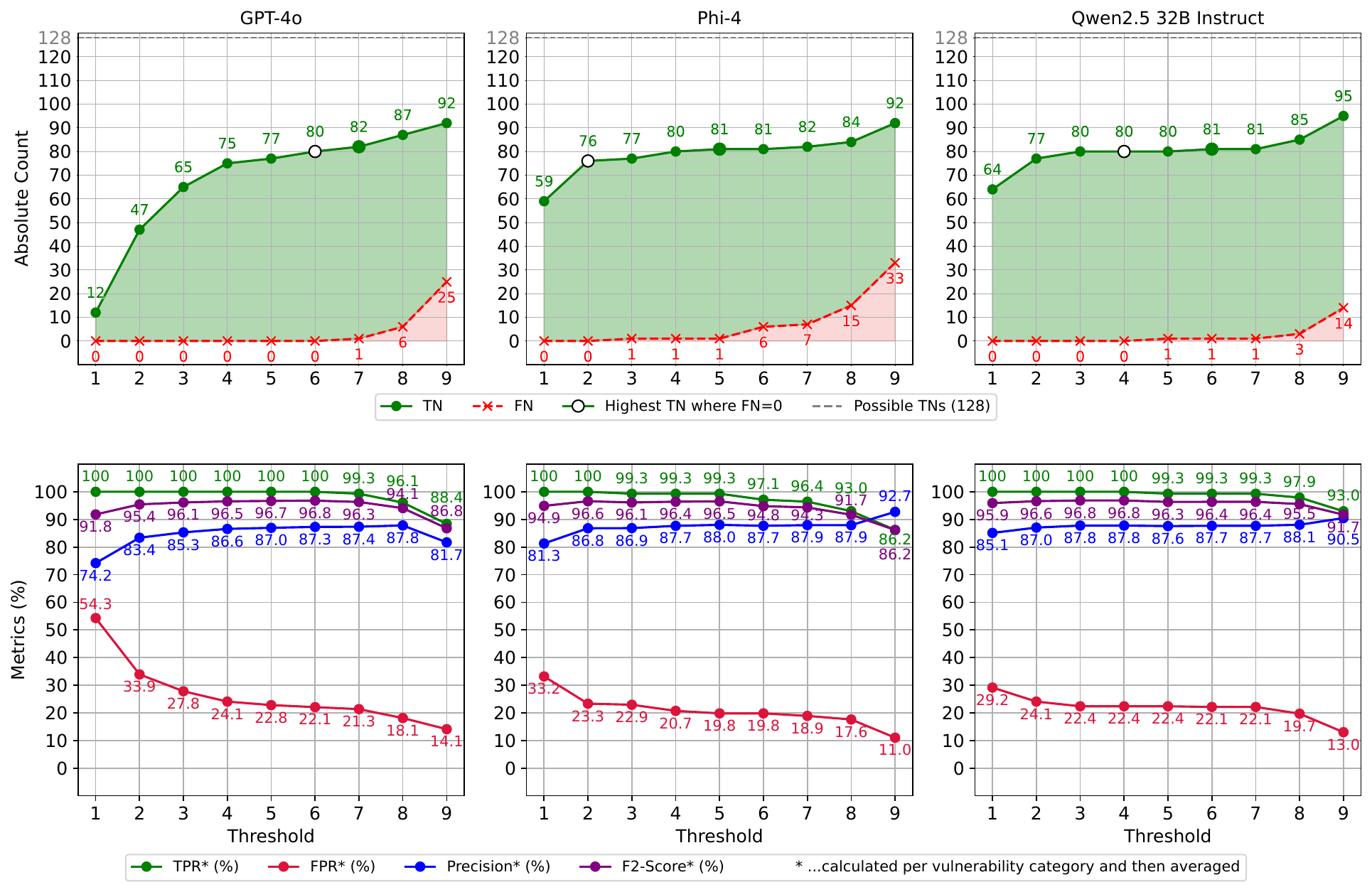}
  \vspace*{-1ex}
  \caption{Self-Consistency (n=5) Results of GPT-4o, Phi-4 and Qwen2.5 32B}
  \label{fig:llms_sc_comparison}
  \vspace*{-1ex}
\end{figure*}
\section{Analysis and Discussion}\label{sec:analysis}
Building upon the experimental setup described in Section~\ref{sec:experiments}, we now present a comprehensive evaluation of the selected \LLMs. 
The corresponding result files, provided in JSON format, are publicly accessible, as detailed in~\cite{wagner_2025_15378450}.
Specifically, we investigate whether they can conservatively detect FPs in the OWASP test split, analyze the effectiveness of their detection capabilities, and explore the extent to which their predictions can be effectively combined (\ie ensemble decision). 
Subsequently, we examine the performance of our framework on real-world security findings and provide a critical outlook on future research directions.

\subsection{Conservative Analysis}
To provide an overview of whether the considered \LLMs are capable of performing conservative analysis, Fig.~\ref{fig:llms_min_thershold} presents the 3-shot CoT classification results of all evaluated \LLMs at the lowest tested threshold (threshold = 1.0). 
Evaluating the \LLM predictions using this user-defined threshold of 1.0 implies that all security findings classified by an \LLM with a returned score $\geq 1.0$ are flagged as vulnerable. 
Conversely, only test cases receiving a decision value below 1.0 (\eg decision = 0.0) are flagged as FPs.
In Fig.~\ref{fig:llms_min_thershold}, the y-axis represents the proportion of genuine weaknesses missed by each \LLM (FNs), while the x-axis illustrates the proportion of FPs correctly identified (TNs).
Our primary objective is thus to maximize the detection rate of TN classifications, while keeping the FNs at zero (conservative analysis).
Fig.~\ref{fig:llms_min_thershold} indicates that six distinct \LLMs are successfully performing the assessment without producing any FNs at the lowest evaluated threshold. 
These \LLMs are \texttt{GPT-4o}, \texttt{Phi-4}, \texttt{Qwen 2.5 32B}, \texttt{Qwen 2.5 72B}, \texttt{Phi-3 Medium 128k}, and \texttt{Qwen 2.5 Coder 7B}, recognizable by the green dotted line at the top. 
Notably, \texttt{Phi-4} emerges as the most effective \LLM for conservative analysis at this threshold, accurately detecting a substantial proportion (approx. 46.1\%; 59 out of 128) of all FPs in our dataset. 
Conversely, although models like \texttt{Llama 3.1 70B} identify a remarkably high proportion of FPs (approx. 86.7\%; 111 out of 128), they do not meet our criteria for conservative analysis, as they simultaneously fail to preserve all genuine weaknesses.
Thus, with regard to \textbf{RQ1}, the findings derived from Fig.~\ref{fig:llms_min_thershold} demonstrate that it is indeed possible to conservatively eliminate FPs from SAST tool reports while preserving all genuine weaknesses. 
Contrary to our initial expectations, some of the largest \LLMs (particularly certain \texttt{Llama} and \texttt{Gemini} variants) did not manage to achieve a conservative detection of FPs.

\subsection{Effectiveness of Conservative Analysis}
Having established that some \LLMs can indeed achieve FP detection without missing genuine weaknesses, we now optimize the effectiveness of this capability.
The y-axis of Fig.~\ref{fig:llms_thresholds_comp} illustrates, for each \LLM, the highest decision threshold at which no FNs occurred (indicating conservative analysis) in our 3-shot CoT setting. 
Additionally, the proportion of FPs each \LLM detects (indicated by the TN ratio) while maintaining a perfect \TPR (no FNs) is shown on the x-axis. The numbers in brackets next to each \LLM's proportion of TN classifications denote the relative improvement compared to the lowest evaluated threshold (threshold = 1.0) presented in Fig.~\ref{fig:llms_min_thershold}. 
\texttt{GPT-4o} emerged as the superior \LLM, achieving the highest proportion of TNs (approx. 59.4\%; 76 out of 128), closely followed by \texttt{Phi-4} (approx. 58.6\%; 75 out of 128) and \texttt{Qwen 2.5 32B} (approx. 57.0\%; 73 out of 128). 
It is important to note that achieving a high TN ratio at lower thresholds is preferable, as \LLMs performing effectively at low thresholds demonstrate greater applicability in real-world scenarios, where the optimal threshold is not known in advance.
\begin{figure*}[t]
  \centering  \includegraphics[width=0.99\linewidth]{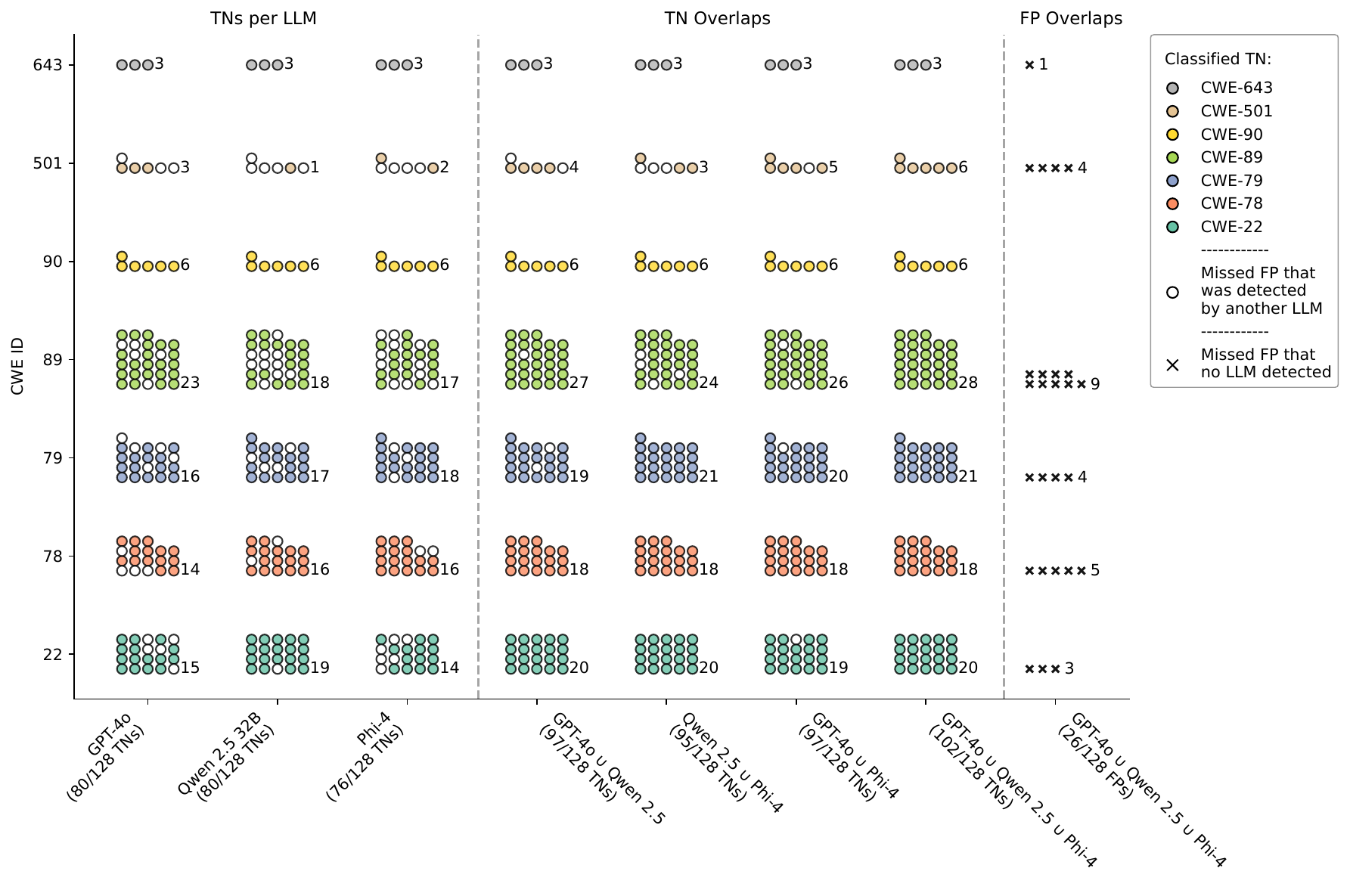}
  \vspace*{-6ex}
  \caption{TN Classification Overlaps of Self-Consistency Results}
  \label{fig:llms_overlap}
\end{figure*}
As described in Sec.~\ref{sec:experiments}, to further improve our conservative analysis and overall effectiveness, we repeated this experiment five times with the three best-performing \LLMs (\texttt{GPT-4o}, \texttt{Phi-4}, and \texttt{Qwen2.5 32B}) and aggregated the results using the SC strategy. 
Fig.~\ref{fig:llms_sc_comparison} visualizes the performance metrics for these \LLMs across nine decision thresholds following the SC approach.
We focus our evaluation on the introduced metrics: $\TPR$, $\FPR$, $\Precision$, and \FtwoScore (all calculated per vulnerability category and then averaged to avoid bias) and the proportion of TNs, representing detected FPs.
The performance metrics are presented in the second row of Fig.~\ref{fig:llms_sc_comparison}, along with the absolute TN and FN counts shown in the first row.
There, a white dot marks the best threshold for each \LLM—that is, the highest decision threshold at which no FNs occurred. 
For \texttt{GPT-4o}, the optimal threshold is 6, yielding a TN count of 80 (62.5\% detected FPs) with zero FNs. 
\texttt{Phi-4} achieves its best performance at threshold 2 with 76 TNs (59.4\% detected FPs), while \texttt{Qwen2.5-32B-Instruct} reaches a maximum TN of 80 (62.5\% detected FPs) at threshold 3.
Although the optimal thresholds vary among the \LLMs, the results depicted in Fig.~\ref{fig:llms_sc_comparison} indicate that SC prompting increased conservative FP detection across all three evaluated \LLMs.
\texttt{GPT-4o}--detecting 80 out of 128 FPs--results in a TN proportion of 62.5\%; this marks an enhancement of 3.1 percentage points (+4 TNs).
With \texttt{Qwen2.5-32B-Instruct} also detecting 80 out of 128 FPs again resulting in a TN proportion of 62.5\% this marks an enhancement of 5.5 percentage points (+7 TNs).
Notably, this high FP detection effectiveness was attained while conducting a conservative analysis. 
As shown in Fig.~\ref{fig:llms_sc_comparison}, the \TPR remains at 100\%, and both \Precision and weighted \FtwoScore exhibit robust performance at the optimal thresholds, reaching up to 96.8\% (\FtwoScore) in \texttt{GPT-4o} and \texttt{Qwen2.5-32B-Instruct}, and 87.8\% for the \Precision.
These results underscore the effectiveness of our conservative FP detection approach via the SC mechanism: not only can a significant portion of FPs be flagged (up to 62.5\%), but this can be achieved without missing any genuine weaknesses. 
This finding addresses \textbf{RQ2} and confirms that our SC mechanism further enhances the FP detection effectiveness in our framework, with multiple \LLMs demonstrating the capability for effective conservative filtering.

\subsection{Combining Results of Multiple Conservative Analyses}
To address \textbf{RQ3}, we further investigated only the top three performing \LLM configurations after applying our SC approach. 
We compared their TN classifications, focusing on uniquely and jointly detected FPs (\ie how much their FP detection capabilities overlap).
Fig.~\ref{fig:llms_overlap} illustrates the overlap of TN classifications across these \LLMs by vulnerability categories (CWE-IDs), revealing both a substantial common core and smaller sets of uniquely detected FPs. 
Crucially, despite considerable overlap, each \LLM identified unique FPs missed by the others, clearly exemplified by CWE-ID 501 (Trust Boundary Violation), where no overlap exists among detected FPs, indicating that the evaluated \LLMs do more than merely replicate one another{}’s decisions.
Given that the \LLMs in this experiment operated at thresholds ensuring conservative analyses, their unique FP detections (TN classifications) could be aggregated. 
Merging the \LLMs{}’ TN classifications at these conservative thresholds increases the overall FP detection rate without compromising the detection of genuine weaknesses.
\texttt{GPT-4o} and \texttt{Qwen2.5-32B-Instruct} each reached up to 80 TNs at their optimal threshold. 
Combining the respective TN sets of our three best-performing \LLMs results in the detection of 102 out of 128 FPs in the dataset, which corresponds to approximately 78.9\% of FPs being detected, all while maintaining a perfect \TPR, with approximate scores of 98.1\% \FtwoScore and 91.3\% \Precision.
Thus, addressing \textbf{RQ3}, the results confirm that the three considered \LLMs have mutually complementary capabilities. 
While agreeing on a core set of FPs, their individual classifications yield additional FPs that collectively raise the effectiveness of our approach, with combined predictions enhancing the relative number of detected FPs from 62.5\% (\texttt{GPT-4o} and \texttt{Qwen2.5-32B-Instruct}) up to 78.9\% on the OWASP test split.

\subsection{Conservative Analysis of Real-World Security Findings}
To prove that our presented framework is valuable in a real-world security context, we used the three best-performing \LLMs from our previous analysis (\texttt{GPT-4o}, \texttt{Qwen2.5-32B-Instruct}, and \texttt{Phi-4}) to assess our real-world dataset (49 TPs and 65 FPs) using the SC approach.
In contrast to the results observed on the OWASP Benchmark test split, all \LLMs performed best at a threshold of 2. \texttt{GPT-4o} identified 24 out of the 65 FPs but missed one genuine weakness and therefore did not achieve a conservative analysis on the real-world dataset. Both \texttt{Qwen2.5-32B-Instruct} and \texttt{Phi-4} successfully performed conservative analyses, with \texttt{Phi-4} correctly filtering out 33.85\% of all FPs (22 out of 65). Moreover, since \texttt{Qwen2.5-32B-Instruct} and \texttt{Phi-4} each include thresholds for conservative analyses, the TN outputs from both \LLMs can be combined. This ensemble detects 25 FPs, representing an overall detection rate of 38.46\% across all FPs, while still preserving every genuine weakness.
Although the achieved FP detection rates are lower than those observed on the OWASP Benchmark, this discrepancy is expected and attributable to the increased complexity and heterogeneity of the real-world dataset.
Specifically, our real-world dataset involved three programming languages, diverse infrastructure file formats, and five independent SAST tools, contrasting with the controlled environment of the OWASP Benchmark where SpotBugs analyzed a Java-only dataset.
Nevertheless, our proposed approach again demonstrated practical applicability by performing a conservative analysis, highlighting its out-of-the-box effectiveness in a realistic software development environment, thereby underscoring its generalizability and value for real-world security assessments.

\subsection{Threats to Validity and Future Work}
Our study has several limitations. First, the \LLM performance strongly depends on selected prompting techniques and contextual information. Minor changes in prompts could significantly alter results. Additionally, most few-shot examples were Java-specific, potentially limiting generalizability. Although we addressed LLM nondeterminism (\eg setting the temperature to zero), minor variability may persist. The chosen thresholds also depend on labeled datasets, making it challenging to guarantee conservative analyses on unlabeled data.
Further, due to the OWASP Benchmark being publicly available on the internet, we cannot know whether it was included in an \LLM{}’s training data.

Future work could examine the specific strengths of individual LLMs across vulnerability categories and further develop ensemble-based strategies for improved FP detection.
Similar, crafting task-specific prompts (\eg for vulnerability categories with high error rates) could positively influence a models performance. Exploring additional advanced prompting methods and evaluating fine-tuned LLMs versus general-purpose models would also provide valuable insights.

\section{Conclusion}\label{sec:conclusion}
This study aims to enhance the effectiveness of security assessments performed with Static Application Security Testing (SAST) tools by drastically reducing manual work. 
We propose a generalized approach leveraging Large Language Models (\LLMs) to automatically detect incorrectly reported weaknesses (false positives, FPs). We present a conservative analysis strategy, aiming to detect FPs while not missing genuine weaknesses, significantly saving valuable expert resources.
Our approach utilizes advanced prompting techniques, particularly Chain-of-Thought and Self-Consistency, without requiring resource-intensive fine-tuning (\ie increasing efficiency). Evaluations conducted on the OWASP Benchmark (v1.2) demonstrated that our best-performing \LLMs (GPT-4o and Qwen2.5-32B-Instruct) conservatively identified 62.5\% of all removable findings, achieving an \FtwoScore of 96.8\% and a \Precision of 87.8\%, without missing any genuine weaknesses. Combining reassessment outputs from multiple LLMs would increase FP detection to approximately 78.9\%, with approximate scores of 98.1\% \FtwoScore and 91.3\% \Precision.
Additionally, we validated our methodology using a complex, heterogeneous real-world dataset, where the best-performing LLM (Phi-4) conservatively detected 33.85\% of FPs, with combined assessments increasing this detection to 38.46\%.

Ultimately, our methodology complements traditional SAST tools, significantly reducing human resource demands, enhancing automation, and establishing a strong baseline for future LLM-based FP detection in security assessments.
Our results suggest that once an LLM is proven to filter conservatively, its performance can be enhanced through threshold optimization, and ensembles of conservatively filtering models will preserve all genuine findings while further boosting FP detection.

\bibliographystyle{IEEEtran}
\bibliography{bibliography}

\begin{thebibliography}{10}
\providecommand{\url}[1]{#1}
\csname url@samestyle\endcsname
\providecommand{\newblock}{\relax}
\providecommand{\bibinfo}[2]{#2}
\providecommand{\BIBentrySTDinterwordspacing}{\spaceskip=0pt\relax}
\providecommand{\BIBentryALTinterwordstretchfactor}{4}
\providecommand{\BIBentryALTinterwordspacing}{\spaceskip=\fontdimen2\font plus
\BIBentryALTinterwordstretchfactor\fontdimen3\font minus \fontdimen4\font\relax}
\providecommand{\BIBforeignlanguage}[2]{{%
\expandafter\ifx\csname l@#1\endcsname\relax
\typeout{** WARNING: IEEEtran.bst: No hyphenation pattern has been}%
\typeout{** loaded for the language `#1'. Using the pattern for}%
\typeout{** the default language instead.}%
\else
\language=\csname l@#1\endcsname
\fi
#2}}
\providecommand{\BIBdecl}{\relax}
\BIBdecl

\bibitem{MitigatingFalsePositiveStaticAnalysisWarnings2023}
Z.~Guo, T.~Tan, S.~Liu, X.~Liu, W.~Lai \emph{et~al.}, ``Mitigating {False} {Positive} {Static} {Analysis} {Warnings}: {Progress}, {Challenges}, and {Opportunities},'' \emph{IEEE TSE}, vol.~49, no.~12, pp. 5154--5188, 2023.

\bibitem{Falsenegativethatoneisgoingtokillyou2024}
A.~S. Ami, K.~Moran, D.~Poshyvanyk, and A.~Nadkarni, ````{False} negative - that one is going to kill you'': Understanding industry perspectives of static analysis based security testing,'' in \emph{IEEE Symposium on Security and Privacy}.\hskip 1em plus 0.5em minus 0.4em\relax IEEE, 2024, pp. 3979--3997.

\bibitem{spa}
A.~M\o{}ller and M.~I. Schwartzbach, ``Static program analysis,'' October 2018, department of Computer Science, Aarhus University, \texttt{http://cs.au.dk/\~{}amoeller/spa/}.

\bibitem{owasp_benchmark}
\BIBentryALTinterwordspacing
{OWASP Foundation}, ``{OWASP Benchmark Project},'' 2016. [Online]. Available: \url{https://web.archive.org/web/20240522054757/https://owasp.org/www-project-benchmark/}
\BIBentrySTDinterwordspacing

\bibitem{koc_learning_2017}
U.~Koc, P.~Saadatpanah, J.~S. Foster, and A.~A. Porter, ``\BIBforeignlanguage{en}{Learning a classifier for false positive error reports emitted by static code analysis tools},'' in \emph{\BIBforeignlanguage{en}{1st {ACM} {SIGPLAN} {International} {Workshop} on {Machine} {Learning} and {Programming} {Languages}}}.\hskip 1em plus 0.5em minus 0.4em\relax ACM, 2017, pp. 35--42.

\bibitem{koc_empirical_2019}
U.~Koc, S.~Wei, J.~S. Foster, M.~Carpuat, and A.~A. Porter, ``An {Empirical} {Assessment} of {Machine} {Learning} {Approaches} for {Triaging} {Reports} of a {Java} {Static} {Analysis} {Tool},'' in \emph{12th {IEEE} {Conference} on {Software} {Testing}, {Validation} and {Verification} ({ICST})}.\hskip 1em plus 0.5em minus 0.4em\relax IEEE, 2019, pp. 288--299.

\bibitem{thapa_transformer-based_2022}
C.~Thapa, S.~I. Jang, M.~E. Ahmed, S.~Camtepe, J.~Pieprzyk, and S.~Nepal, ``\BIBforeignlanguage{en}{Transformer-{Based} {Language} {Models} for {Software} {Vulnerability} {Detection}},'' in \emph{\BIBforeignlanguage{en}{Proceedings of the 38th {Annual} {Computer} {Security} {Applications} {Conference}}}.\hskip 1em plus 0.5em minus 0.4em\relax USA: ACM, 2022, pp. 481--496.

\bibitem{bakhshandeh_using_2023}
A.~Bakhshandeh, A.~Keramatfar, A.~Norouzi, and M.~M. Chekidehkhoun, ``Using {ChatGPT} as a {Static} {Application} {Security} {Testing} {Tool},'' 2023, arXiv:2308.14434 [cs].

\bibitem{jensen_software_2024}
R.~I.~T. Jensen, V.~Tawosi, and S.~Alamir, ``Software {Vulnerability} and {Functionality} {Assessment} using {LLMs},'' 2024, arXiv:2403.08429 [cs].

\bibitem{wen_automatically_2024}
C.~Wen, Y.~Cai, B.~Zhang, J.~Su, Z.~Xu \emph{et~al.}, ``\BIBforeignlanguage{en}{Automatically {Inspecting} {Thousands} of {Static} {Bug} {Warnings} with {Large} {Language} {Model}: {How} {Far} {Are} {We}?}'' \emph{\BIBforeignlanguage{en}{ACM Transactions on Knowledge Discovery from Data}}, vol.~18, no.~7, pp. 1--34, 2024.

\bibitem{li_assisting_2023}
H.~Li, Y.~Hao, Y.~Zhai, and Z.~Qian, ``\BIBforeignlanguage{en}{Assisting {Static} {Analysis} with {Large} {Language} {Models}: {A} {ChatGPT} {Experiment}},'' in \emph{\BIBforeignlanguage{en}{31st {ACM} {Joint} {European} {Software} {Engineering} {Conference} and {Symposium} on the {Foundations} of {Software} {Engineering}}}.\hskip 1em plus 0.5em minus 0.4em\relax ACM, 2023, pp. 2107--2111.

\bibitem{zhou_comparison_2024}
X.~Zhou, D.-M. Tran, T.~Le-Cong, T.~Zhang, I.~C. Irsan \emph{et~al.}, ``Comparison of {Static} {Application} {Security} {Testing} {Tools} and {Large} {Language} {Models} for {Repo}-level {Vulnerability} {Detection},'' Jul. 2024, arXiv:2407.16235 [cs].

\bibitem{tamberg_harnessing_2025}
K.~Tamberg and H.~Bahsi, ``Harnessing {Large} {Language} {Models} for {Software} {Vulnerability} {Detection}: {A} {Comprehensive} {Benchmarking} {Study},'' \emph{IEEE Access}, vol.~13, pp. 29\,698--29\,717, 2025, arXiv:2405.15614 [cs].

\bibitem{nori_can_2023}
H.~Nori, Y.~T. Lee, S.~Zhang, D.~Carignan, R.~Edgar \emph{et~al.}, ``Can {Generalist} {Foundation} {Models} {Outcompete} {Special}-{Purpose} {Tuning}? {Case} {Study} in {Medicine},'' 2023, arXiv:2311.16452 [cs].

\bibitem{wei_chain--thought_2023}
J.~Wei, X.~Wang, D.~Schuurmans, M.~Bosma, and B.~m.~o. Ichter, ``Chain-of-{Thought} {Prompting} {Elicits} {Reasoning} in {Large} {Language} {Models},'' 2023, arXiv:2201.11903 [cs].

\bibitem{kojima_large_2023}
T.~Kojima, S.~S. Gu, M.~Reid, Y.~Matsuo, and Y.~Iwasawa, ``Large {Language} {Models} are {Zero}-{Shot} {Reasoners},'' 2023, arXiv:2205.11916.

\bibitem{wang_self-consistency_2023}
X.~Wang, J.~Wei, D.~Schuurmans, Q.~Le, E.~Chi \emph{et~al.}, ``Self-{Consistency} {Improves} {Chain} of {Thought} {Reasoning} in {Language} {Models},'' 2023, arXiv:2203.11171 [cs].

\bibitem{wagner_2025_15378450}
\BIBentryALTinterwordspacing
J.~Wagner, ``Towards efficient complementary security analysis using large language models,'' May 2025. [Online]. Available: \url{https://doi.org/10.5281/zenodo.15378450}
\BIBentrySTDinterwordspacing

\bibitem{10.1145/3606367}
\BIBentryALTinterwordspacing
P.~Christen, D.~J. Hand, and N.~Kirielle, ``A review of the f-measure: Its history, properties, criticism, and alternatives,'' \emph{ACM Comput. Surv.}, vol.~56, no.~3, Oct. 2023. [Online]. Available: \url{https://doi.org/10.1145/3606367}
\BIBentrySTDinterwordspacing

\bibitem{openai_gpt-4_2024}
OpenAI, J.~Achiam, S.~Adler, S.~Agarwal, L.~Ahmad \emph{et~al.}, ``{GPT}-4 {Technical} {Report},'' 2024, arXiv:2303.08774 [cs].

\bibitem{gpt_4o}
\BIBentryALTinterwordspacing
{OpenAI Team}, ``Hello {GPT-4o},'' May 2024, {Blog} Post. [Online]. Available: \url{https://web.archive.org/web/20240815014626/https://openai.com/index/hello-gpt-4o/}
\BIBentrySTDinterwordspacing

\bibitem{dubey_llama_2024}
A.~Dubey, A.~Jauhri, A.~Pandey, A.~Kadian, A.~Al-Dahle \emph{et~al.}, ``The {Llama} 3 {Herd} of {Models},'' 2024, arXiv:2407.21783 [cs].

\bibitem{gemini_team_gemini_2024}
{Gemini Team}, P.~Georgiev, V.~I. Lei, R.~Burnell, L.~Bai \emph{et~al.}, ``Gemini 1.5: {Unlocking} multimodal understanding across millions of tokens of context,'' 2024, arXiv:2403.05530 [cs].

\bibitem{abdin_phi-3_2024}
M.~Abdin, S.~A. Jacobs, A.~A. Awan, J.~Aneja, A.~Awadallah \emph{et~al.}, ``Phi-3 {Technical} {Report}: {A} {Highly} {Capable} {Language} {Model} {Locally} on {Your} {Phone},'' 2024, arXiv:2404.14219 [cs].

\bibitem{abdin_phi-4_2024}
M.~Abdin, J.~Aneja, H.~Behl, S.~Bubeck, R.~Eldan \emph{et~al.}, ``Phi-4 {Technical} {Report},'' Dec. 2024, arXiv:2412.08905 [cs].

\bibitem{bai_qwen_2023}
J.~Bai, S.~Bai, Y.~Chu, Z.~Cui, K.~Dang \emph{et~al.}, ``Qwen {Technical} {Report},'' Sep. 2023, arXiv:2309.16609 [cs].

\bibitem{yang_qwen2_2024}
A.~Yang, B.~Yang, B.~Hui, B.~Zheng, B.~Yu, C.~Zhou \emph{et~al.}, ``Qwen2 {Technical} {Report},'' Sep. 2024, arXiv:2407.10671 [cs].

\bibitem{qwen_qwen25_2025}
Qwen, A.~Yang, B.~Yang, B.~Zhang, B.~Hui \emph{et~al.}, ``Qwen2.5 {Technical} {Report},'' Jan. 2025, arXiv:2412.15115 [cs].

\bibitem{hui_qwen25-coder_2024}
B.~Hui, J.~Yang, Z.~Cui, J.~Yang, D.~Liu \emph{et~al.}, ``Qwen2.5-{Coder} {Technical} {Report},'' Nov. 2024, arXiv:2409.12186 [cs].

\end{thebibliography}

\end{document}